\documentclass[prd,twocolumn,aps]{revtex4}
\usepackage{amsmath}
\usepackage{graphicx}

\def\R{{\cal R}}

\def\vu{^{(5)\!}}
\def\be{\begin{equation}}
\def\ee{\end{equation}}
\def\bea{\begin{eqnarray}}
\def\eea{\end{eqnarray}}

\begin{document}

\title{Observational constraints on braneworld inflation:\\
the effect of a Gauss-Bonnet term}

\author{Shinji Tsujikawa$^1$, M. Sami$^2$ and Roy Maartens$^3$
}

\address{~}
\address{$^1$ Department of Physics, Gunma National College of
Technology, Gunma 371-8530, Japan }

\address{$^2$ IUCAA, Post Bag 4, Ganeshkhind,
Pune 411 007, India }

\address{$^3$ Institute of Cosmology \& Gravitation,
University of Portsmouth, Portsmouth PO1 2EG, UK}

\date{\today}

\begin{abstract}

High-energy modifications to general relativity introduce changes
to the perturbations generated during inflation, and the latest
high-precision cosmological data can be used to place constraints
on such modified inflation models. Recently it was shown that
Randall-Sundrum type braneworld inflation leads to tighter
constraints on quadratic and quartic potentials than in general
relativity. We investigate how this changes with a Gauss-Bonnet
correction term, which can be motivated by string theory.
Randall-Sundrum models preserve the standard consistency relation
between the tensor spectral index and the tensor-to-scalar ratio.
The Gauss-Bonnet term breaks this relation, and also modifies the
dynamics and perturbation amplitudes at high energies. We find
that the Gauss-Bonnet term tends to soften the Randall-Sundrum
constraints. The observational compatibility of the quadratic
potential is strongly improved. For a broad range of energy
scales, the quartic potential is rescued from marginal rejection.
Steep inflation driven by an exponential potential is excluded in
the Randall-Sundrum case, but the Gauss-Bonnet term leads to
marginal compatibility for sufficient e-folds.

\end{abstract}

\maketitle \vskip 1pc

\section{Introduction}

The WMAP, SDSS and other data of high-precision cosmology open up
the possibility of testing and constraining models of the universe
(see e.g. Ref.~\cite{SDSS}). The standard model based on general
relativity and inflation passes all current observational tests
with suitable proportions of dark energy and dark matter, and
without the need for departures from adiabaticity, gaussianity or
scale-invariance in the primordial perturbations. This gives
strong support to the slow-roll inflationary scenario, while at
the same time initiating a period in which the inflationary
parameters can be constrained by observations. Indeed, the viable
slow-roll parameter space is dramatically
reduced~\cite{lealid,GRconst}.

It is not possible to reconstruct the functional form of the
inflationary potential from observations. However, improved
constraints may be derived if a general form of the potential is
assumed, and if appropriate limits are placed on the number $N$ of
e-folds of inflation after horizon-crossing. Some inflationary
potentials are under pressure from the data, with the chaotic
quartic potential ($V\propto\phi^4$) marginally ruled out at the
2$\sigma$ level~\cite{lealid}. Constraints on slow-roll
inflationary potentials follow from the properties of their
primordial perturbation spectra, and in particular from the
degeneracy between scalar and tensor perturbations. This
degeneracy is expressed via the standard consistency relation
$R=-8n_{\rm T}$ (discussed below). The degeneracy arises since a
tensor contribution increases power on large scales, requiring a
compensating increase of small-scale scalar power, and thus a more
blue spectrum $n_{\rm S}>1$.

These recent advances in observational constraints on slow-roll
potentials may be extended to the case where Einstein's general
relativity is modified at inflationary energy scales. If the true
fundamental scale of gravity is lower than the effective
four-dimensional Planck scale $M_4\sim 10^{16}~$TeV, as in
higher-dimensional gravity theories, then significant
modifications to general relativity may arise during inflation.
The simplest phenomenological models describing such a scenario
are the five-dimensional Randall-Sundrum type braneworld
cosmologies (see e.g. Ref.~\cite{review} for recent reviews).
These models are motivated by ideas from string theory and
M~theory. The observable universe is a four-dimensional ``brane"
surface embedded in a five-dimensional anti de Sitter ``bulk"
spacetime, with standard model fields trapped on the brane, while
gravity propagates in the bulk. Although the fifth dimension is
infinite, the curvature of the bulk acts to confine the graviton
near the brane at low energies, so that general relativity is
recovered. At high energies in the early universe, graviton
localization fails and the Friedmann equation is modified. The
amplitudes and spectra of perturbations generated during inflation
are also modified~\cite{Maartens,Langlois}.

Remarkably, despite the modifications to both scalar and tensor
primordial perturbations, the consistency relation $R=-8n_{\rm T}$
is unchanged in this braneworld scenario~\cite{Huey}. (We may
expect some change from higher-order expansions in slow-roll
parameters~\cite{RL}, but this modification is too small to be
detected in current observations, and it is also not clear that
bulk perturbations may be neglected at higher order.) As a
consequence, observational constraints on the primordial
perturbation spectra cannot distinguish between the general
relativistic and braneworld models~\cite{lidtay}. Braneworld
effects on the evolution of perturbations after horizon re-entry
will leave signatures on the cosmic microwave background (CMB),
but the techniques for calculating these signatures are still
under development~\cite{branecmb}. If constraints are only
available from the primordial perturbation spectra, then specific
braneworld parameters cannot be constrained -- but the constraints
on particular slow-roll potentials are modified. For example, the
perturbations produced by the quadratic potential
($V\propto\phi^2$) are further from scale-invariance than in the
standard cosmology, while in the quartic case they are more or
less unchanged~\cite{LS,Shinji}. Thus both of the simplest chaotic
potentials are under strong observational pressure in the simplest
braneworld scenario.

The Randall-Sundrum braneworld cosmology is based on the
five-dimensional Einstein-Hilbert action. At high energies, it is
expected that this action will acquire quantum corrections. String
theory and holography indicate that higher-order curvature
invariants will arise in the action at perturbative level. In five
dimensions, the Gauss-bonnet curvature invariant has special
properties: it is the unique combination that leads to
second-order field equations linear in the second derivatives, and
it is ghost-free (see e.g. Refs.~\cite{CD,MT,LN}). A natural
question is whether the Gauss-Bonnet correction to the
Einstein-Hilbert action will disrupt the degeneracy between tensor
and scalar perturbations. The answer is that the standard
consistency relation is broken by the Gauss-Bonnet
term~\cite{Dufaux}.

However, the breaking of degeneracy is ``mild", and it turns out,
as we shall show, that the likelihood values are almost identical
to those in the standard and Randall-Sundrum cases. Thus the
introduction of a Gauss-Bonnet term does not lead to braneworld
signatures in current observations of the primordial perturbation
spectra. However, it does lead to interesting changes in the
constraints on inflationary potentials. The quadratic potential
moves inside the 1$\sigma$ bound for a range of inflationary
energy scales even for $N=50$, so that the simplest chaotic
inflation model is ``rescued" by the Gauss-Bonnet correction. The
quartic potential is also removed from marginal rejection, and is
inside the 2$\sigma$ bound for a range of energy scales even when
$N=50$. The exponential potential, which can drive ``steep"
inflation in Randall-Sundrum models~\cite{steep}, is excluded by
observations in the Randall-Sundrum case. The Gauss-Bonnet
correction again rescues this potential, moving it inside the
2$\sigma$ bound for a range of energy scales when $N\gtrsim 55$.

\section{Inflation in a Gauss-Bonnet Braneworld}

For a 5D bulk with Einstein-Gauss-Bonnet gravity, containing a 4D
brane, the gravitational action is
 \bea
{\cal S} &=& \frac{1}{2\kappa_5^2} \int d^5x \sqrt{-\,\vu g}
\left[-2\Lambda_5+ {\cal R} \right. \nonumber\\
&&\left.~{}+\alpha\, \left({\cal R}^2-4 {\cal R}_{ab}{\cal
R}^{ab}+ {\cal R}_{abcd}{\cal R}^{abcd}\right) \right] \nonumber\\
&&~{} - \int_{\rm brane} d^4x\, \sqrt{-g}\, \sigma\,,
\label{action}
 \eea
where $x^a=(x^\mu,y)$, with $y$ the extra-dimensional coordinate,
$g_{ab}=\,\vu g_{ab}-n_an_b$ is the induced metric, with $n^a$ the
unit normal to the brane, $\sigma\, (>0)$ is the brane tension,
and $\Lambda_5\,(<0)$ is the bulk cosmological constant. The
fundamental energy scale of gravity is the 5D scale $M_5$, where
$\kappa_5^2=8\pi/M_5^3$. The Planck scale $M_4$ is an effective
scale, describing gravity on the brane at low energies, and
typically $M_4 \gg M_5$.

The Gauss-Bonnet (GB) term may be thought of as the lowest-order
stringy correction to the 5D Einstein-Hilbert action, with
coupling constant $\alpha\geq 0$. In this case,
$\alpha|\R^2|\ll|\R|$, so that $\alpha \ll \ell^2$, where $\ell$
is the bulk curvature scale, $|\R|\sim \ell^{-2}$. In terms of the
extra-dimensional energy scale $\mu\equiv \ell^{-1}$,
 \be
\beta\equiv 4\alpha\mu^2 \ll 1\,. \label{al}
 \ee
For an anti de Sitter bulk, $\Lambda_5=-3\mu^2(2-\beta)$. The
Randall-Sundrum (RS) type models are recovered for $\beta=0$.

Imposing $Z_2$ symmetry across a Friedmann brane in an anti de
Sitter bulk, we get the Friedmann equation~\cite{CD,LN}
\bea \label{hubble} H^2 &=& {\mu^2\over
\beta}\left[(1-\beta)\cosh\left({2\chi\over3}
\right)-1\right]\,,\label{mfe}\\
\label{chi} \kappa_5^2(\rho+\sigma) &=& 2\mu\left[{{2(1-\beta)^3}
\over {\beta} }\right]^{1/2} \sinh\chi\,,
 \eea
where $H$ is the Hubble rate, and $\chi$ is a dimensionless
measure of the energy density $\rho$ on the brane. Eliminating
$\chi$ leads to
\be \kappa_5^2(\rho+\sigma) =
2\mu\sqrt{1+{H^2\over\mu^2}}\left[3-\beta +2\beta
{H^2\over\mu^2}\right]. \label{mf}
 \ee
In order to recover General Relativity (GR) at low energies, the
effective 4D Newton constant is~\cite{CD,MT}
\begin{equation}
\label{k4k5} \kappa_4^2= \frac{\mu} {1+\beta}\, \kappa_5^2\,.
\end{equation}
The brane tension is fine-tuned by achieving a zero cosmological
constant on the brane~\cite{MT}:
\begin{equation}
\kappa_5^2\sigma = 2\mu(3-\beta)\,. \label{sig'}
\end{equation}

By Eqs.~(\ref{hubble}) and (\ref{chi}) there is a characteristic
GB energy scale,
 \be
m_\beta=\left[{8\mu^2(1-\beta)^3\over
\beta\kappa_5^4}\right]^{1/8},
 \ee
such that the GB high energy regime ($\sinh\chi \gg 1$)
corresponds to $\rho \gg m_\beta^4$. Since the GB term is a
correction to the RS action, we must have $m_\beta^4$ greater than
the RS energy scale $\sigma$; this requires~\cite{Dufaux} $\beta<
0.152$, which is consistent with Eq.~(\ref{al}). Expanding
Eq.~(\ref{mfe}) in $\chi$, we find three regimes for the dynamical
history of the brane universe~\cite{Dufaux}
 \bea
\rho\gg m_\beta^4~& \Rightarrow &~ H^2\approx \left[
{\mu^2\kappa_5^2 \over 4\beta}\, \rho \right]^{\!2/3}({\rm
GB~regime}),
\label{vhe}\\
m_\beta^4 \gg \rho\gg\sigma~& \Rightarrow &~ H^2\approx
{\kappa_4^2
\over 6\sigma}\, \rho^{2}~~~({\rm RS~regime}),\label{he}\\
\rho\ll\sigma~ & \Rightarrow &~ H^2\approx {\kappa_4^2 \over 3}\,
\rho~~~({\rm GR~regime}). \label{gr} \eea

We are mainly interested in inflation which occurs in the GB
regime, Eq.~(\ref{vhe}), and in the intermediate regime between
Eqs.~(\ref{vhe}) and (\ref{he}). These are the regimes where the
GB correction is crucial. We assume a 4D inflaton field $\phi$
with potential $V(\phi)$ on the brane, which satisfies the
Klein-Gordon equation,
\bea \ddot{\phi}+3H\dot{\phi}+V_\phi(\phi)=0\,. \label{KG} \eea
During slow-roll inflation, $|\ddot{\phi}| \ll |3H\dot{\phi}|$ and
$\dot{\phi}^2 \ll V(\phi)$. Since $V\approx\rho \gg \sigma$,
Eq.~(\ref{chi}) gives $V=m_\beta^4 \sinh\chi$. Making use of
Eq.~(\ref{hubble}), we find that the slow-roll parameters,
$\epsilon \equiv -\dot{H}/H^2$ and $\eta \equiv
V_{\phi \phi}/3H^2$, are given by
 \bea
\label{epsilon} {\epsilon \over \epsilon_{\rm RS}} &=&\!
\frac{2(1-\beta)^4\sinh {2\over3}\chi \tanh \chi \, \sinh^2
\chi}{9(1+\beta) (3-\beta)[ (1-\beta) \cosh {2\over3}\chi -1 ]^2}\!,\\
\label{eta} {\eta \over \eta_{\rm RS}} &=&\!
\frac{2(1-\beta)^3\sinh^2 \chi}{3 (1+\beta)(3-\beta)[(1-\beta)
\cosh {2\over3}\chi -1]} ,
 \eea
where $\epsilon_{\rm RS}$ and $\eta_{\rm RS}$ are the RS slow-roll
parameters~\cite{Maartens},
\begin{equation}
\epsilon_{\rm RS}=\frac{2\sigma}{\kappa_4^2}
~\frac{V_\phi^2}{V^3}\,,~~ \eta_{\rm RS}= \frac{2\sigma}
{\kappa_4^2} \frac{V_{\phi \phi}}{V^2} \,. \label{etaRS}
\end{equation}
In the limit $\chi \ll 1$ we have $\epsilon \to \epsilon_{\rm RS}$
and $\eta \to \eta_{\rm RS}$. The number of e-folds, $N \equiv
\int Hdt$, is
\begin{equation}
\label{efold} N \approx 3\int_{\chi_e}^{\chi_N}
\frac{H^2}{V_{\chi}} \left(\frac{{\rm d}\phi}{{\rm
d}\chi}\right)^2 {\rm d}\chi\,,
\end{equation}
where $\chi_e$ is evaluated at the end of inflation
($\epsilon=1$), and $\chi_N$ is evaluated when cosmological scales
leave the horizon. This value of $\chi$ is subject to a quantum
gravity upper limit: in order to consistently apply semi-classical
analysis of inflationary perturbations, we require $V < M_5^4 $,
which yields the constraint~\cite{Dufaux}
\begin{equation}
\label{const} \sinh\chi < \left({M_5\over m_\beta}\right)^4\,.
\end{equation}

We consider simple monomial potentials corresponding to chaotic
large-field inflation:
\begin{equation}
\label{po} V(\phi)=V_0\phi^p\,.
\end{equation}
We are mainly interested in the cases $p=2,4$ and $p \to \infty$
(exponential potential). The slow-roll parameters become
\begin{eqnarray}
\label{epsilon2} \epsilon &=& \left[\frac{4\sigma p^2V_0^{2/p}}
{27\kappa_4^2 m_\beta^{4+8/p}}\right] f(\chi)\,, \\
\label{eta2} \eta &=& \left[\frac{4\sigma p(p-1)V_0^{2/p}}
{9\kappa_4^2 m_\beta^{4+8/p}}\right] g(\chi)\,,
\end{eqnarray}
where
 \bea
f(\chi) &=& \frac{\sinh {2\over3}\chi \, \tanh \chi \, (\sinh
\chi)^{1-2/p}}{\left[ \cosh {2\over3}\chi -1 \right]^2} \,,\\
g(\chi) & =& \frac{(\sinh\chi)^{1-2/p}}{\cosh {2\over3}\chi -1}\,,
 \eea
and we used Eq.~(\ref{al}). In the high-energy regime $\sinh\chi
\gg 1$, we have $f(\chi), g(\chi) \propto \exp[(p-6)\chi/3p]$.
Therefore for $p<6$ both slow-roll parameters grow with the
decrease of $\chi$. On the other hand, in the case $p>6$,
including steep inflation, the situation is opposite, as pointed
out in Ref.~\cite{LN}. In the regime $\chi \ll 1$ one has
$\epsilon, \eta \propto \chi^{-1-2/p}$, so that both slow-roll
parameters increase with decreasing $\chi$ for all positive values
of $p$.

\begin{figure}
\begin{center}
\includegraphics[height=3.3in,width=3.5in]{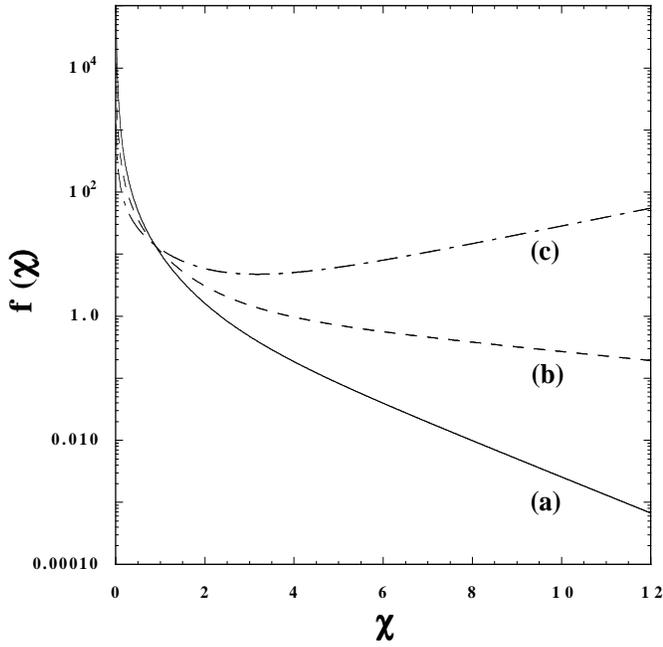}
\caption{ The dependence of the slow-roll parameter $\epsilon$ on
$\chi$ [see Eq.~(\ref{epsilon2})] for (a)~$p=2$, (b)~$p=4$ and
(c)~$p \to \infty$.  }\label{slowpara}\end{center}

\end{figure}

In Fig.~\ref{slowpara} we show $f(\chi)$ for $p=2,4, \infty$. The
behaviour of $g(\chi)$ is qualitatively similar. When $p \to
\infty$, the slow-roll parameter $\epsilon$ gets smaller with the
decrease of $\chi$ in the GB regime ($\chi \gtrsim 3$), after
which it begins to grow in the RS regime. Therefore inflation
comes to an end in the RS regime as argued in Ref.~\cite{LN}. On
the other hand $\epsilon$ always increases with the decrease of
$\chi$ for $p=2$ and $p=4$, as seen in Fig.~\ref{slowpara}. This
means that inflation can end either in the RS or GB regime,
depending on the amplitude of the coefficient terms in
Eqs.~(\ref{epsilon2}) and (\ref{eta2}). In order to estimate the
maximum values of $|\epsilon|$ and $|\eta|$ for cosmologically
relevant scales, it is sufficient to consider the case where
inflation terminates in the RS regime ($\chi_e \ll 1$). This is
because one obtains smaller values of $|\epsilon|$ and $|\eta|$
when inflation ends in the GB regime.

The number of e-folds, Eq.~(\ref{efold}), is
\bea \label{efold2} N &=& \frac{3\mu^2}{\beta p^2V_0^{2/p}
m_\beta^{4-8/p}} \int_{\chi_e}^{\chi_N} \frac{\left[\cosh
{2\over3}\chi-1\right] \cosh \chi} {(\sinh \chi)^{2-2/p}}d\chi
\nonumber\\
&\equiv& \frac{3\mu^2}{\beta p^2V_0^{2/p} m_\beta^{4-8/p}} \Big[
I(\chi) \Big]_{\chi_e}^{\chi_N}\,. \eea
One can find an explicit form of the integral $I(\chi)$ for
particular values of $p$, as we see later.

Let us consider the case where inflation ends in the RS regime
($\chi_e \ll 1$). By Eq.~(\ref{efold2}), $I(\chi) \sim
2p\chi^{(2+p)/p}/9(2+p)$ for $\chi \ll 1$ (setting the integration
constant to zero), so that
\begin{equation}
\label{efold3} N =\frac{3\mu^2}{\beta p^2V_0^{2/p}
m_\beta^{4-8/p}} \left[I(\chi_N)-\frac{2p}{9(2+p)}
\chi_e^{1+2/p}\right]\,.
\end{equation}
By Eq.~(\ref{epsilon}) we obtain for $\chi \ll 1$
\begin{eqnarray}
\label{epslow} \epsilon=\left[\frac{2\sigma p^2V_0^{2/p}}
{\kappa_4^2 m_\beta^{4+8/p}}\right] \chi^{-1-2/p}\,,
\end{eqnarray}
so that the value of $\chi$ at the end of inflation is
\begin{eqnarray}
\label{chie} \chi_e=\frac{V_0}{m_\beta}\left(\frac{2\sigma p^2}
{\kappa_4^2V_0}\right)^{p/(p+2)}\,.
\end{eqnarray}
Combining these results gives the following relation
\begin{eqnarray}
\label{rela} \frac{2\sigma p^2V_0^{2/p}} {\kappa_4^2
m_\beta^{4+8/p}} =\frac{9(2+p)}{2\left[N(2+p)+p\right]}
I(\chi_N)\,.
\end{eqnarray}
Then the slow-roll parameters, Eqs.~(\ref{epsilon2}) and
(\ref{eta2}), may be written for $\chi_e\ll1$ in the form
\begin{eqnarray}
\label{epsilonf} \epsilon &=&
\frac{(2+p)I(\chi_N)}{3\left[N(2+p)+p\right]} \,f(\chi_N) \,, \\
\label{etaf} \eta &=&
\frac{(p-1)(p+2)I(\chi_N)}{p\left[N(2+p)+p\right]} \, g(\chi_N)\,.
\end{eqnarray}
These expressions agree with Ref.~\cite{LN} in the special case $p
\to \infty$.

\begin{figure}
\begin{center}
\includegraphics[height=3.5in,width=3.5in]{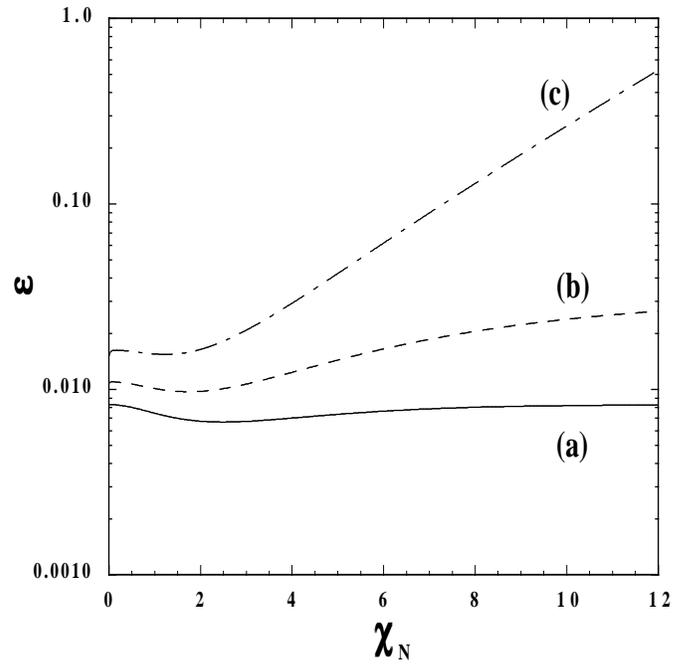}
\caption{ The dependence of $\epsilon$ on $\chi_N$, for $N=60$ and
(a)~$p=2$, (b)~$p=4$, (c)~$p \to \infty$.
}\label{feps}\end{center}
\end{figure}

\begin{figure}
\begin{center}
\includegraphics[height=3.5in,width=3.5in]{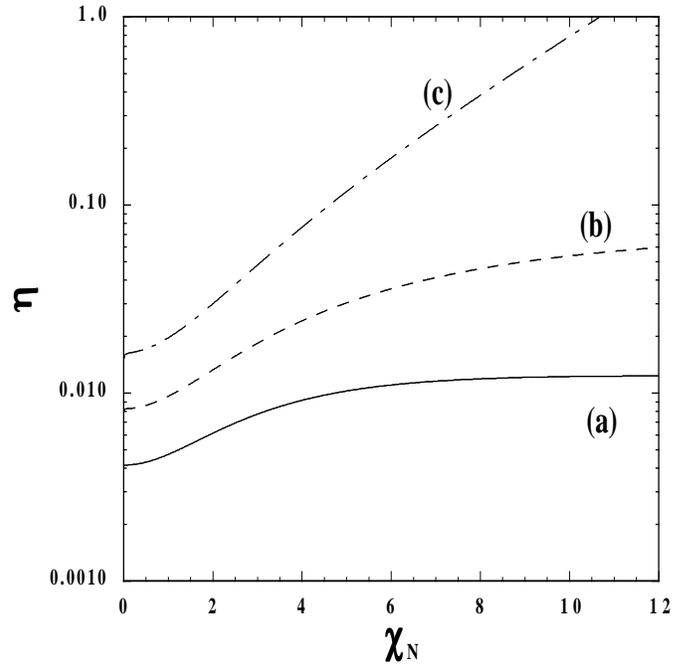}
\caption{ The dependence of $\eta$ on $\chi_N$, for $N=60$ and
(a)~$p=2$, (b)~$p=4$, (c)~$p \to \infty$.
}\label{feta}\end{center}
\end{figure}

When $p=2$ the integral in Eq.~(\ref{efold2}) can be evaluated:
\begin{equation}
\label{func} I(\chi)=\frac32\left[\!\cosh{2\over3}\chi-1 - {\ln}\!
\left(\!{1\over3}+{2\over3}\cosh{2\over3}
\chi\!\right)\!\right]\!,
\end{equation}
where the integration constant has been chosen so that $I(\chi)
\sim \chi^2/9$ for $\chi \ll 1$. When $p=4$ we do not have an
analytical form, but the integral can be found numerically. In the
limit $p \to \infty$ the integral is given by~\cite{LN}
\begin{eqnarray}
\label{rela2} I(\chi)=\frac{2}{\sqrt{3}}\tan^{-1}\! \left(\frac{2}
{\sqrt{3}}\sinh {\chi\over3}\!\right)\!+\frac{1-\cosh
{2\over3}\chi} {\sinh \chi}\,,
\end{eqnarray}
which behaves as $I(\chi) \sim 2\chi/9$ for $\chi \ll 1$.

In Figs.~\ref{feps} and \ref{feta} we plot the slow-roll
parameters as functions of $\chi_N$ for $p=2,4,\infty$. We
integrated Eq.~(\ref{efold2}) numerically and checked that this
shows excellent agreement with the analytical results for $p=2$
and $p \to \infty$. The GB ($\sinh\chi_N \gg 1$) and RS ($\chi_N
\ll 1$) regimes are connected by an intermediate regime of a
combination of GB and RS inflation. Figure~\ref{feps} shows that
$\epsilon$ has a minimum in this region ($1.5 \lesssim \chi_N
\lesssim 3$). This is an important feature for estimating the
ratio of tensor to scalar perturbations, as we see in the next
section.

The parameter $\eta$ is a monotonically increasing function. It is
also evident from Figs.~\ref{feps} and \ref{feta} that both
slow-roll parameters increase with an increase in $p$. Note,
however, that we can have $\epsilon,\eta<1$ even in the limit $p
\to \infty$, which means that inflation can be realized for a
steep exponential potential, as in the RS case. If we choose
smaller values of $N$ than shown in Figs.~\ref{feps} and
\ref{feta}, the slow-roll parameters tend to increase for a given
value of $\chi_N$. The qualitative behaviour of $\epsilon$ and
$\eta$ is similar to that illustrated in Figs.~\ref{feps} and
\ref{feta} for cosmologically relevant scales ($50 \lesssim N
\lesssim 70$).

\section{Perturbation spectra}

In this section we evaluate the spectra of perturbations produced
during slow-roll inflation and place constraints on the GB
braneworld inflationary models using a compilation of the latest
observational data.

The amplitude of scalar perturbations is given by~\cite{Dufaux}
\begin{eqnarray}
A_{\rm S}^2 &=& \frac{9H^6}{2\pi^2V_\phi^2}
=\left[{\kappa_4^6V^3 \over 6\pi^2V_\phi^2}
\right]G_\beta^2(H/\mu) \,,
 \eea
where the term in square brackets is the standard amplitude, and
the GB braneworld modification is given by
 \be \label{gg}
G^2_\beta(x)=\left[{3(1+\beta) x^2 \over 2\sqrt{1+x^2}(3-\beta
+2\beta x^2)+2(\beta-3)}\right]^{3},
 \ee
where $x\equiv H/\mu$ is a dimensionless measure of energy scale.
The RS amplification factor~\cite{Maartens} is recovered when
$\beta=0$. For the monomial potential, we find that
 \be
\!\!A_{\rm S}^2=
\frac{3^{3-1/p}\kappa_4^{2-2/p}\sigma^{-1+1/p}\mu^{4+2/p}}
{2^{3-2/p}\pi^2p^2V_0^{2/p}\beta^{2+1/p}} \frac{\left[\cosh
{2\over3}\chi -1 \right]^3}{(\sinh \chi)^{2-2/p}}
\,.\label{scalar}
 \ee
The scalar spectral index is
\begin{eqnarray}
\label{nS} n_{\rm S}-1 \equiv \frac{{\rm d\, ln}\,A_{\rm
S}^2}{{\rm d\,ln}\,k} \Biggr|_{k=aH} =-6\epsilon+2\eta\,.
\end{eqnarray}

The amplitude of tensor perturbations is~\cite{Dufaux}
\begin{equation}
A_{\rm T}^2=\left[\kappa_4^2 \,\frac{H^2}{4\pi^2}\right]
F_{\beta}^2(H/\mu)\,, \label{tensor}
\end{equation}
where the standard expression is in square brackets, and the GB
braneworld modification is defined by
\begin{equation}
F_\beta^{-2}(x)=\sqrt{1+x^2}-\left(\frac{1-\beta}{1+\beta}
\right)x^2 \sinh^{-1}\frac{1}{x}\,. \label{F}
\end{equation}
The RS amplification factor~\cite{Langlois} is recovered when
$\beta=0$. The tensor spectral index $n_{\rm T}$ is
\begin{eqnarray}
\label{nT} &&\!\!\!n_{\rm T}=\frac{{\rm d\, ln}\,A_{\rm T}^2}
{{\rm
d\,ln}\,k}\Biggr|_{k=aH}\nonumber\\
&&\!\!\!=\! -2\epsilon\left[\!1-\!
\frac{x^2F_\beta^2(x)\{\!1-(1-\beta) \sqrt{1+x^2} \sinh^{-1}
x^{-1}\}} {(1+\beta)\sqrt{1+x^2}}\!
\right]\!. \nonumber \\
\end{eqnarray}

The tensor-to-scalar ratio is given by
\begin{eqnarray}
\label{ratio2} R \equiv 16\frac{A_{\rm T}^2}{A_{\rm S}^2}=
-8Q\,n_{\rm T}\,,
\end{eqnarray}
where~\cite{Dufaux}
\begin{eqnarray}
 Q&=& \frac{(1-\beta)\cosh \chi} {[1+2(1-\beta)\sinh^2
{1\over3 }\chi]\cosh {1\over3 }\chi} \nonumber\\ &=&
\frac{1+\beta+2\beta x^2}{1+\beta+\beta x^2}\,.\label{Q}
\end{eqnarray}
The standard form of the consistency relation, $Q=1$, $R=-8n_{\rm
T}$, holds in the absence of the Gauss-Bonnet coupling ($\beta =
0$), i.e., in both the GR and RS cases. When $\beta=0$ we have
$n_{\rm T} \to -3\epsilon$ for $x \to \infty$ (RS model) and
$n_{\rm T} \to -2\epsilon$ for $x \to 0$ (GR case). When
$\beta>0$, the degeneracy factor $Q\to 2$ in the GB limit ($x \to
\infty$), thus yielding the consistency relation $R=-16n_{\rm T}$.
In this case one has $n_{\rm T} \to -\epsilon$ for $x \to \infty$.
In summary:
\begin{eqnarray}
\label{con3} & &n_{\rm T}=-\epsilon\,,~~
R=-16n_{\rm T}=16\epsilon~~~({\rm GB~~limit})\,, \\
\label{RSli}
& &n_{\rm T}=-3\epsilon\,,~~R=-8n_T=24\epsilon~~~({\rm RS~~limit})\,, \\
& &n_{\rm T}=-2\epsilon\,,~~R=-8n_T=16\epsilon~~~({\rm
GR~~limit})\,.
\end{eqnarray}
The behaviour of $n_{\rm T}$ is illustrated in Fig.~\ref{nt}.

\begin{figure}
\begin{center}
\includegraphics[height=3.3in,width=3.5in]{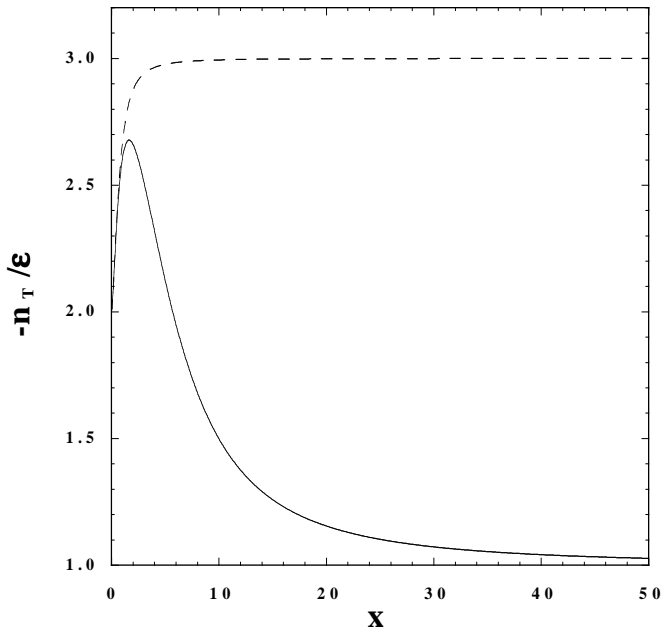}
\caption{The behaviour of $-n_{\rm T}/\epsilon$ as a function of
the energy scale of inflation, $x=H/\mu$. The dashed curve is
$\beta=0$, the solid curve is for $\beta=10^{-2}$. }\label{nt}
\end{center}
\end{figure}

The change of the consistency relation in the GB case can lead to
modification of the likelihood values of inflationary model
parameters. Recently a likelihood analysis in terms of the RS
inflationary parameters has been performed for the consistency
relation $R=-8n_{\rm T}$, using the latest observational
data~\cite{Shinji} (updating Ref.~\cite{LS}). This likelihood
analysis is the same as in GR since the consistency relation is
unchanged. The same applies in generalized Einstein theories,
including the 4-dimensional dilaton gravity and scalar-tensor
theories~\cite{TG}. The situation is different in the presence of
the Gauss-Bonnet coupling.

In order to determine the difference we carried out the likelihood
analysis by using a new consistency relation $R=-16n_{\rm T}$ in
the GB limit. Since the runnings of scalar and tensor
perturbations are consistent with zero~\cite{lealid}, we varied
three inflationary parameters $n_{\rm S}$, $R$ and $A_{\rm S}^2$
together with four cosmological parameters by assuming a flat
$\Lambda{\rm CDM}$ universe. We ran the Cosmological Monte Carlo
(CosmoMC) code together with the CAMB program~\cite{antony1}. In
addition to the dataset from WMAP~\cite{WMAP}, the 2dF~\cite{2dF}
and SDSS~\cite{SDSS} galaxy redshift surveys, we implement the
band powers on smaller scales from VSA~\cite{VSA}, CBI~\cite{CBI}
and ACBAR~\cite{ACBAR}. The 2D posterior constraints in terms of
$n_{\rm S}$ and $R$ are illustrated in Fig.~\ref{likelihood}. This
result is similar to those in Refs.~\cite{Shinji,TG}, which were
derived using the standard consistency relation $R=-8n_{\rm T}$. A
small difference appears for the $2\sigma$ contour bound in the
region $n_{\rm S} >1$, but this is not important to constrain the
inflationary models with monomial potential, Eq.~(\ref{po}). Thus
we can safely use the contour bounds based on $R=-16n_{\rm T}$ to
constrain RS inflationary models with GB correction.

\begin{figure}
\begin{center}
\includegraphics[height=3.5in,width=3.5in]{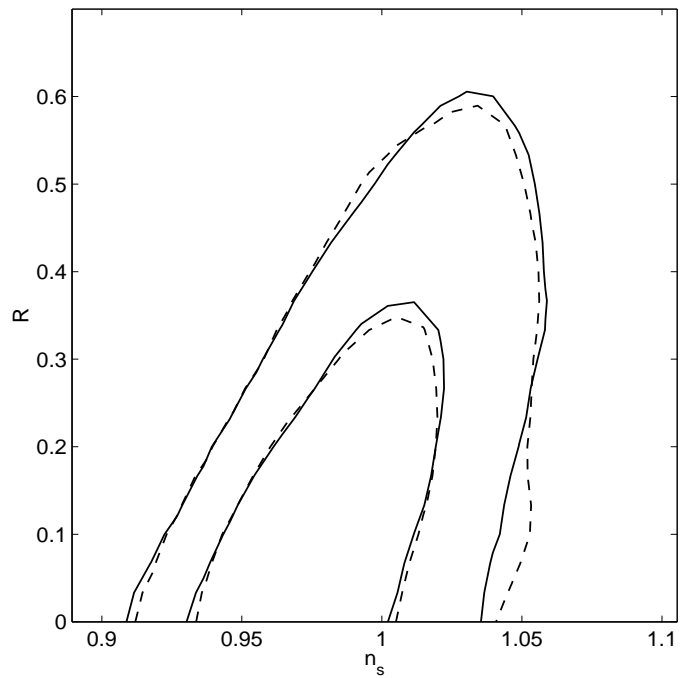}
\caption{ 2D posterior constraints in the $n_{\rm S}$-$R$ plane
with the $1\sigma$ and $2\sigma$ contour bounds. The solid curve
is obtained by using the consistency relation $R=-16n_{\rm T}$ in
the GB limit, whereas the dashed curve corresponds to $R=-8n_{\rm
T}$ in the RS/GR cases. }\label{likelihood}\end{center}
\end{figure}

For the monomial potential, Eq.~(\ref{rela}) gives
 \be \label{sigal}
\sigma^{(p-2)/2p}\beta^{(p+2)/2p} ={2^{-1+2/p}(2+p) \mu^{1+2/p}
\kappa_4^{1-2/p} I(\chi_N) \over 3^{(2-3p)/2p}p^2V_0^{2/p}
[N(2+p)+p]}\!.
 \ee
On the other hand, applying the COBE normalization $A_{\rm
S}^2=5\times 10^{-9}$ in Eq.~(\ref{scalar}) yields
 \be
\label{COBEcon} \sigma^{1-1/p}\beta^{2+1/p}= \frac{3^{3-1/p}10^8
\mu^{4+2/p} \kappa_4^{2-2/p} [\cosh {2\over3}\chi_N -1 ]^3}
{2^{2-2/p}\pi^2 p^2V_0^{2/p}(\sinh \chi_N)^{2-2/p}}\!.
 \ee
As pointed out in Ref.~\cite{LN}, one can determine $\sigma$ and
$\beta$ for given values of $V_0, N, \chi_N$ and $p$ by using
Eqs.~(\ref{sigal}) and (\ref{COBEcon}). Combining these equations,
we obtain
\begin{eqnarray}
\label{sigal2} \frac{\beta^3\sigma }{64\kappa_4^2\mu^6} =\frac{3^3
10^{16}}{(4\pi)^4}
\left[\!\frac{N(2+p)+p}{(2+p)I(\chi_N)}\!\right]^2\! \frac{[\cosh
{2\over3}\chi_N -1]^6} {(\sinh \chi_N)^{4-4/p}}\!.
\end{eqnarray}
This relation is illustrated in Fig.~\ref{sigalcon}. We also plot
the quantum gravity limit derived from Eq.~(\ref{const}), which
may be rewritten as
 \be
48 \sinh^6\chi< (8\pi)^8 \left({\beta^3\sigma \over
64\kappa_4^2\mu^6} \right), \label{qgl2}
 \ee
where we have used Eqs.~(\ref{al}), (\ref{k4k5}) and (\ref{sig'}).
We find the upper limits $\chi_N<11.65$ for $p=2$, $\chi_N<11.22$
for $p=4$ and $\chi_N<10.31$ for $p \to \infty$. This information
is important to place limits on the values of $n_{\rm S}$ and $R$.

\begin{figure}
\begin{center}
\includegraphics[height=3.5in,width=3.5in]{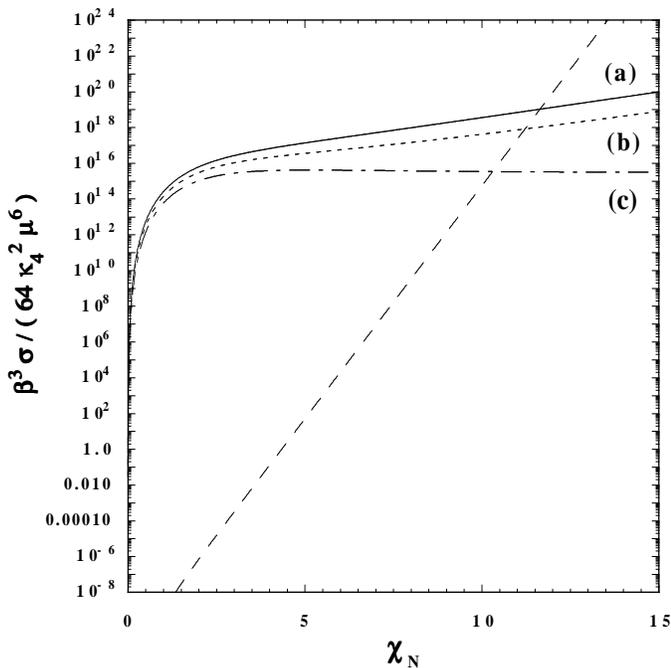}
\caption{ The dependence of the quantity $\beta^3\sigma
/64\kappa_4^2\mu^6$ on $\chi_N$ for (a)~$p=2$, (b)~$p=4$, (c)~$p
\to \infty$. We also show the quantum gravity limit curve,
Eq.~(\ref{qgl2}). The allowed region is above and to the left of
this curve. }\label{sigalcon}\end{center}

\end{figure}

By Eqs.~(\ref{epsilonf}), (\ref{etaf}), (\ref{nS}), (\ref{nT}),
(\ref{ratio2}) and (\ref{Q}) one can evaluate the spectral index
$n_{\rm S}$ and the ratio $R$ as a function of $\chi_N$ for given
values of $N$ and $p$. In Fig.~\ref{nsfig} we plot $n_{\rm S}$ for
$N=60$. When $p=2$ and $p=4$ the spectral index has a maximum at
$\chi_N\sim 4$. This is associated with the fact that the
slow-roll parameter $\epsilon$ has a minimum in the intermediate
region (see Fig.~\ref{feps}). Although $\eta$ is an increasing
function of $\chi_N$, the effect of $\epsilon$ is more important
in $n_{\rm S}$. The spectral index $n_{\rm S}$ decreases for
larger $\chi_N$ in the GB region because of the increase of
$\epsilon$. In the case of the exponential potential ($p \to
\infty$), $n_{\rm S}$ is a monotonically increasing function. It
was shown in Ref.~\cite{LN} that one gets an exactly
scale-invariant spectrum, $n_{\rm S}=1$, for the background
evolution characterized by $H^2 \propto \rho^{2/3}$. Therefore the
spectral index approaches $n_{\rm S}=1$ in the GB limit
($\sinh\chi_N \gg 1$).

Figure~\ref{ratiofig} shows the behaviour of the tensor to scalar
ratio $R$, determined by Eqs.~(\ref{nT}) and (\ref{Q}). We find
that the exact results for $\chi_N \gtrsim 4$ are well
approximated by the relation $R=16\epsilon$ which holds in the GB
limit. The difference appears in the RS regime, where the
consistency relation is $R=24\epsilon$. The ratio $R$ has a
minimum around $\chi_N\sim 2$, depending on the value of $p$. This
is again linked to the fact that the slow-roll parameter
$\epsilon$ has a minimum in this intermediate region. Beyond this
region, $R$ is an increasing function of $\chi_N$ (limited by the
quantum gravity constraint). When $p \to \infty$ the value of $R$
becomes too large for the observational contour bounds, as we see
later.

The values of $n_{\rm S}$ and $R$ in the RS limit
are~\cite{Shinji}
\begin{eqnarray}
\label{nRRS} n_{\rm S}-1=-\frac{2(2p+1)}{N(p+2)}\,,~~~
R=\frac{24p}{N(p+2)}\,.
\end{eqnarray}
These agree with our results shown in Figs.~\ref{nsfig} and
\ref{ratiofig} in the RS regime. The perturbations in the RS case
tend to be further from the point $(n_{\rm S}=1, R=0)$ relative to
the standard GR case for small exponents $p$~\cite{LS}. This is
mainly due to the fact that $\epsilon$ is larger in the RS case. A
similar feature arises in the intermediate regime between RS and
GB. Since the background equation changes from $H^2 \propto
\rho^2$ (RS) to $H^2 \propto \rho^{2/3}$ (GB) with the growth of
$\chi_N$, there is an intermediate region in which the background
is characterized by $H^2 \propto \rho$. This leads to the minimum
of $R$ seen in Fig.~\ref{ratiofig}.

\begin{figure}
\begin{center}
\includegraphics[height=3.5in,width=3.5in]{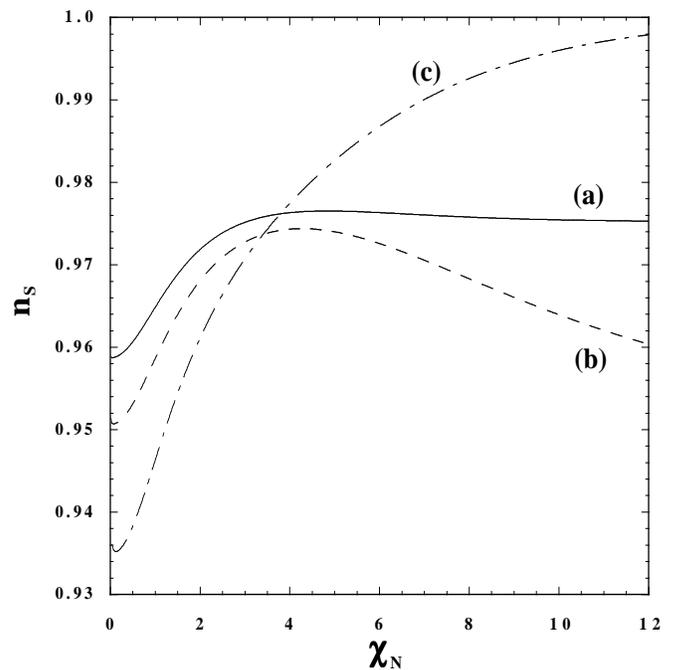}
\caption{ The spectral index $n_{\rm S}$ as a function of $\chi_N$
when $N=60$, for (a)~$p=2$, (b)~$p=4$, (c)~$p \to \infty$.
}\label{nsfig}\end{center}

\end{figure}

\begin{figure}
\begin{center}
\includegraphics[height=3.5in,width=3.5in]{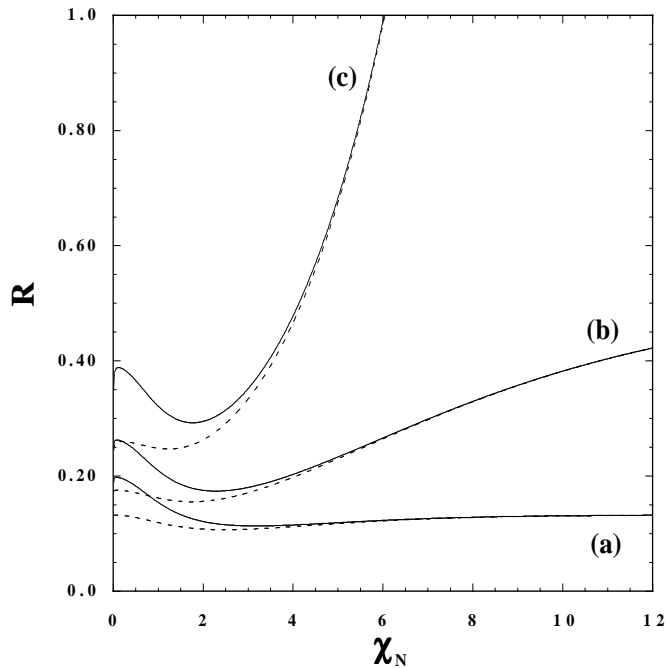}
\caption{ The tensor to scalar ratio $R$ as a function of $\chi_N$
when $N=60$, for (a)~$p=2$, (b)~$p=4$, (c)~$p \to \infty$ (solid
curves). The dotted curves are obtained from $R=16\epsilon$, which
only holds in the GB limit. }\label{ratiofig}\end{center}

\end{figure}

\section{Observational constraints}

By plotting the theoretical values of $n_{\rm S}$ and $R$ in the
2D posterior observational contour bounds of
Fig.~\ref{likelihood}, we can constrain the GB inflationary models
with monomial potential. The parametric theoretical curves
$(n_{\rm S}(\chi_N),R(\chi_N))$ are plotted for values of $\chi_N$  up to
the quantum gravity limit $\chi_N \approx (11.65, 11.22, 10.31)$,
for $p=(2,4,\infty)$ respectively.
\\

\noindent{ $\bullet ~{V\propto \phi^2}$}

In the RS case, the quadratic potential ($p=2$) is inside the
$2\sigma$ contour for $N=50$, and marginally on the $1\sigma$
contour bound for $N=60$, as seen from the left edge point of the
curve (a) in Figs.~\ref{nandRN50} and \ref{nandRN60}. This RS
inflationary model is thus observationally allowed for $N\gtrsim
50$~\cite{LS,Shinji}. As seen in Fig.~\ref{nsfig}, $n_{\rm S}$ for
$\beta > 0$ gets closer to scale-invariance than the RS limiting
value. In addition, Fig.~\ref{ratiofig} indicates that $R$ is
smaller than in the RS case. Therefore the theoretical points tend
to move deeper inside the observationally allowed region, as shown
in Figs.~\ref{nandRN50} and \ref{nandRN60}. The RS limiting case
is marginally allowed for $N=50$, but the GB coupling moves the
points inside the $1\sigma$ contour bound. Thus the quadratic
potential is observationally more favoured when
the GB term is present.\\

\noindent{ $\bullet ~{V\propto \phi^4}$}

The quartic potential ($p=4$) is under strong observational
pressure in both RS~\cite{LS,Shinji} and GR~\cite{GRconst,lealid}
cases. In fact in the RS case it is outside the $2\sigma$ contour
bound for $N<60$, as seen in the left edge point of the curve (b)
in Fig.~\ref{nandRN50}. The GB term leads to an increase of
$n_{\rm S}$ and decrease of $R$ in the intermediate regime between
RS and GB. As in the case $p=2$, this property improves the
observational compatibility of the GB brane-world. However in the
extreme GB limit for the quartic potential, the opposite happens,
and the theoretical points move back into the RS region.

In summary, the GB term rescues the quartic potential as an
observationally allowed model, for inflationary energy scales in
the intermediate regime.\\

\noindent{ $\bullet ~{V=V_0\exp (\xi\kappa_4\phi)}$}

Here $\xi$ is a dimensionless constant. The discussion in Secs.~II
and III can be applied by taking the limit $p \to \infty$ and
making the replacement $p^2V_0^{2/p}~\mapsto~\xi^2\kappa_4^2$. In
the RS case the tensor to scalar ratio is too large to be
compatible with observations~\cite{LS}, as seen in left edge point
of curve (c) in Figs.~\ref{nandRN50} and \ref{nandRN60}. We
require $N\gtrsim 90$ in order for the RS point to be inside the
$2\sigma$ bound, but this is beyond cosmologically relevant
scales.

The GB term works in a similar way to the previous cases. There is
a minimum value of $R$ in the intermediate regime between RS and
GB. For $N=50$ this effect is not sufficient: as seen in
Fig.~\ref{nandRN50}, the theoretical curve is closer but still
outside the $2\sigma$ contour bound. However some of the
theoretical points move inside the $2\sigma$ bound for $N \gtrsim
55$, as shown in Fig.~\ref{nandRN60}. Then steep inflation is not
necessarily ruled out, provided that $\chi_N$ corresponds to an
intermediate energy scale, between RS and GB. Note that $R$
rapidly grows with $\chi_N$ in the GB regime.
Figure~\ref{ratiofig} shows that for $N=60$, $R=1.00$ at
$\chi_N\approx 6$. Steep inflation in the GB regime is disfavoured
due to the large tensor to scalar ratio, in spite of the fact that
the spectral index $n_{\rm S}$ approaches 1.

\begin{figure}
\begin{center}
\includegraphics[height=3.5in,width=3.5in]{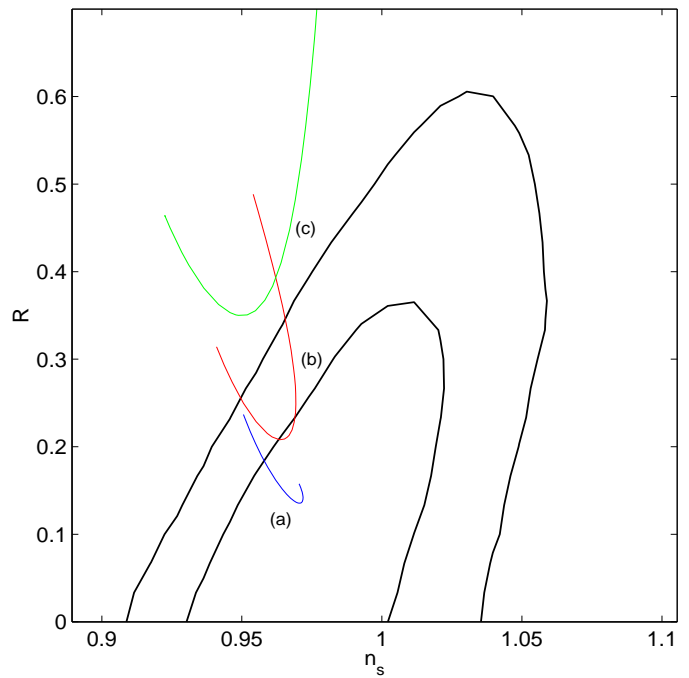}
\caption{ Theoretical prediction of Gauss-Bonnet monomial
inflation models for $N=50$, together with the $1\sigma$ and
$2\sigma$ observational contour bounds. The left edge point of the
theoretical curves is the RS limit, $\chi_N\ll1$, and $\chi_N$
increases along the curves up to the quantum gravity limit. The
curves are shown for (a)~$p=2$, (b)~$p=4$, (c)~$p \to \infty$.
}\label{nandRN50}\end{center}

\end{figure}

\begin{figure}
\begin{center}
\includegraphics[height=3.5in,width=3.5in]{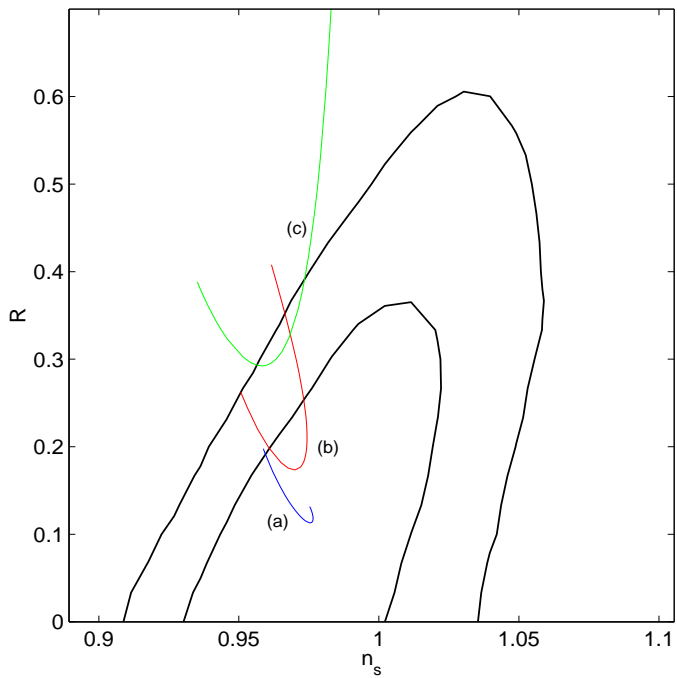}
\caption{ As in Fig.~\ref{nandRN50}, with $N=60$.
}\label{nandRN60}\end{center}

\end{figure}

\section{Conclusions}

In this paper we have studied the observational constraints on
Gauss-Bonnet brane-world inflation, using the latest observational
datasets. The presence of the GB term as a correction to the RS
model modifies the standard consistency relation, $R=-8n_{\rm T}$.
We carried out a likelihood analysis in terms of inflationary and
cosmological parameters by using the GB consistency relation,
$R=-16n_{\rm T}$, which holds in the high-energy GB limit. The
likelihood result is almost identical to the one in the standard
case, which means that this can be safely used even in an
intermediate regime between GB and RS.

We have considered monomial potentials, Eq.~(\ref{po}), including
the exponential potential as a limiting case. The slow-roll
parameters $\epsilon$ and $\eta$ are evaluated as functions of the
dimensionless energy scale $\chi_N$ and the e-folds $N$ [see
Eqs.~(\ref{epsilonf}) and (\ref{etaf})]. An important feature we
found is that $\epsilon$ has a minimum in the intermediate region
between the GB and RS limits, which particularly affects the ratio
$R$ of tensor to scalar perturbations. We evaluated $R$ and the
spectral index $n_{\rm S}$ of scalar perturbations as functions of
$\chi_N$. Then we plotted these theoretical values together with
the 2D observational constraints in the $n_{\rm S}$-$R$ plane. As
seen in Fig.~\ref{ratiofig}, $R$ has a minimum for intermediate
energies, irrespective of the values of $p$. In addition $n_{\rm
S}$ is a growing function for $\chi_N \lesssim 2-3$. Therefore in
this region we obtain smaller $R$ and larger $n_{\rm S}$ than in
the RS case. This feature of GB brane-world inflation improves the
observational compatibility of the 3 potentials considered,
relative to the RS case.

For the quadratic potential ($p=2$), we found that the theoretical
points move inside the $1\sigma$ contour bound for $N=50$, while
the RS point is marginally inside of the $2\sigma$ bound (see
Fig.~\ref{nandRN50}). This means that the quadratic potential is
favoured observationally even for rather low e-folds in the
presence of the GB term.

It is known that the quartic potential ($p=4$) is under strong
observational pressure both in the RS and GR cases, but this
situation is improved by a GB correction term.
Figure~\ref{nandRN50} shows that there exists a region of
parameter space which is inside the $2\sigma$ bound even for
$N=50$.

The exponential potential ($p \to \infty$) is ruled out in the RS
case. However the GB term allows smaller values of $R$, thereby
giving rise to some regions which are inside the $2\sigma$ bound
for $N \gtrsim 55$ (see Fig.~\ref{nandRN60}). This is an
intriguing possibility to save the brane-world steep inflation
models (e.g., Ref.~\cite{SV}). It may also be of interest to
extend our analysis to the case in which the system is described
by the Dirac-Born-Infeld-type action on the brane, since steep
potentials of the type $V(\phi) \propto \exp(m^2\phi^2)$ appear in
such a theory~\cite{Reza}.

\[\]
{\bf Acknowledgments:} We are grateful to Sam Leach for providing
the latest SDSS code and Antony Lewis and David Parkinson for
technical help in numerics. The work of RM is supported by PPARC.



\begin{thebibliography}{99}

\bibitem{SDSS}
D.~N.~Spergel {\it et al.},
Astrophys.\ J.\ Suppl.\  {\bf 148}, 175 (2003)
[arXiv:astro-ph/0302209]; M.~Tegmark {\it et al.}  [SDSS
Collaboration],
Phys.\ Rev.\ D{\bf 69}, 103501 (2004) [arXiv:astro-ph/0310723].

\bibitem{GRconst}
H.~V.~Peiris {\it et al.},
Astrophys.\ J.\ Suppl.\  {\bf 148}, 213 (2003); V.~Barger,
H.~S.~Lee, and D.~Marfatia,
Phys.\ Lett.\ B{\bf 565}, 33 (2003) [arXiv:hep-ph/0302150];
W.~H.~Kinney, E.~W.~Kolb, A.~Melchiorri, and A.~Riotto,
arXiv:hep-ph/0305130.

\bibitem{lealid}
S.~M.~Leach and A.~R.~Liddle, Phys. Rev. D{\bf 68}, 123508 (2003)
[arXiv:astro-ph/0306305].

\bibitem{review}
P.~Brax and C.~van de Bruck, Class. Quantum Grav. {\bf 20}, R201
(2003) [arXiv:hep-th/0303095]; R.~Maartens, Liv. Rev. Rel., to
appear (2004) [arXiv:gr-qc/0312059]; P.~Brax, C.~van de Bruck, and
A-C. Davis, arXiv:hep-th/0404011.

\bibitem{Maartens}
R.~Maartens, D.~Wands, B.~A.~Bassett and I.~Heard,
Phys.\ Rev.\ D{\bf 62}, 041301 (2000) [arXiv:hep-ph/9912464].

\bibitem{Langlois}
D.~Langlois, R.~Maartens, and D.~Wands,
Phys.\ Lett.\ B{\bf 489}, 259 (2000) [arXiv:hep-th/0006007].

\bibitem{Huey}
G.~Huey and J.~E.~Lidsey,
Phys.\ Lett.\ B{\bf 514}, 217 (2001) [arXiv:astro-ph/0104006];
D.~Seery and A.~Taylor,
arXiv:astro-ph/0309512.

\bibitem{RL}
G.~Calcagni,
JCAP {\bf 0311}, 009 (2003) [arXiv:hep-ph/0310304]; ibid.,
arXiv:hep-ph/0312246; E.~Ramirez and A.~R.~Liddle,
Phys.\ Rev.\ D {\bf 69}, 083522 (2004) [arXiv:astro-ph/0309608].

\bibitem{lidtay}
A.~R.~Liddle and A.~N. Taylor, Phys. Rev. D{\bf 65}, 041301 (2002)
[arXiv:astro-ph/0109412].

\bibitem{branecmb}
K.~Koyama, Phys. Rev. Lett. {\bf 91}, 221301 (2003)
[arXiv:astro-ph/0303108]; C.~S.~Rhodes, C.~van de Bruck, Ph.~Brax,
A. C. Davis,  Phys. Rev. D{\bf 68}, 083511 (2003)
[arXiv:astro-ph/0306343].

\bibitem{LS}
A.~R.~Liddle and A.~J.~Smith,
Phys.\ Rev.\ D{\bf 68}, 061301 (2003) [arXiv:astro-ph/0307017].

\bibitem{Shinji}
S.~Tsujikawa and A.~R.~Liddle,
JCAP {\bf 03}, 001 (2004) [arXiv:astro-ph/0312162].

\bibitem{CD}
C.~Charmousis and J.~F.~Dufaux,
Class.\ Quant.\ Grav.\  {\bf 19}, 4671 (2002)
[arXiv:hep-th/0202107]; S.~C.~Davis,
Phys.\ Rev.\ D{\bf 67}, 024030 (2003) [arXiv:hep-th/0208205];
P.~Binetruy, C.~Charmousis, S.~C.~Davis and J.~F.~Dufaux,
Phys.\ Lett.\ B{\bf 544}, 183 (2002) [arXiv:hep-th/0206089].

\bibitem{MT}
K.~Maeda and T.~Torii,
Phys.\ Rev.\ D{\bf 69}, 024002 (2004) [arXiv:hep-th/0309152].

\bibitem{LN}
J.~E.~Lidsey and N.~J.~Nunes,
Phys.\ Rev.\ D{\bf 67}, 103510 (2003) [arXiv:astro-ph/0303168]

\bibitem{Dufaux}
J.~F.~Dufaux, J.~E.~Lidsey, R.~Maartens and M.~Sami,
arXiv:hep-th/0404161.

\bibitem{steep}
E.~J.~Copeland, A.~R.~Liddle, and J.~E.~Lidsey,
Phys.\ Rev.\ D{\bf 64}, 023509 (2001) [arXiv:astro-ph/0006421].

\bibitem{TG}
S.~Tsujikawa and B.~Gumjudpai, 
Phys.\ Rev.\ D {\bf 69}, 123523 (2004)
[arXiv:astro-ph/0402185].

\bibitem{antony1}
A.~Lewis, A.~Challinor, and A.~Lasenby,
Astrophys.\ J.\  {\bf 538}, 473 (2000) [arXiv:astro-ph/9911177];
A.~Lewis and S.~Bridle,
Phys.\ Rev.\ D{\bf 66}, 103511 (2002) [arXiv:astro-ph/0205436];
see also http://camb.info/.

\bibitem{WMAP}
http://lambda.gsfc.nasa.gov/

\bibitem{2dF}
W.~J.~Percival {\it et al.},
Mon. Not. Roy. Astron. Soc. {\bf 327}, 1297 (2001)
[arXiv:astro-ph/0105252].

\bibitem{VSA}
K.~Grainge {\it et al.}, Mon. Not. Roy. Astron. Soc. {\bf 341},
L23 (2003) [arXiv:astro-ph/0212495].

\bibitem{CBI}
T.~Pearson {\it et al.}, Astrophys. J. {\bf 591}, 556 (2003)
[arXiv:astro-ph/0205388].

\bibitem{ACBAR}
C.~L.~Kuo {\it et al.}, Astrophys. J. {\bf 600}, 32 (2004)
[arXiv:astro-ph/0212289].

\bibitem{SV}
V.~Sahni, M.~Sami and T.~Souradeep,
Phys.\ Rev.\ D {\bf 65}, 023518 (2002) [arXiv:gr-qc/0105121];
M.~Sami and V.~Sahni,
arXiv:hep-th/0402086.

\bibitem{Reza}
M.~R.~Garousi,
JHEP {\bf 0312}, 036 (2003) [arXiv:hep-th/0307197]; M.~R.~Garousi,
M.~Sami and S.~Tsujikawa,
arXiv:hep-th/0402075; ibid., arXiv:hep-th/0405012.



\end{thebibliography}
\end{document}